# Routing centralization across domains via SDN: A model and emulation framework for BGP evolution

Vasileios Kotronis[a,*], Adrian Gämperli[a], Xenofontas Dimitropoulos[b]

[a] *ETH Zurich, Switzerland*
[b] *Foundation of Research and Technology Hellas (FORTH), Greece*

## ABSTRACT

The Border Gateway Protocol (BGP) was designed almost three decades ago and has many limitations relating to its fully distributed nature, policy enforcement capabilities, scalability, security and complexity. For example, the control plane can take several minutes to converge after a routing change; this may be unacceptable for real-time network services. Despite many research proposals for incremental improvements and clean-slate redesigns of how inter-domain routing should work, BGP is likely one of the most ossified protocols of the Internet architecture and it has not retrofitted the proposed ideas. In this work, we propose a radical, incrementally deployable Internet routing paradigm in which the control plane of multiple networks is logically centralized. This follows the Software Defined Networking (SDN) paradigm, although at the inter-domain level involving multiple Autonomous Systems (AS). Multi-domain SDN centralization can be realized by outsourcing routing functions to an external contractor, which provides inter-domain routing services facilitated through a multi-AS network controller. The proposed model promises to become a vehicle for evolving BGP and uses the bird's eye view over several networks to benefit aspects of inter-domain routing, such as convergence properties, policy conflict resolution, inter-domain troubleshooting, and collaborative security. In addition to the proposed paradigm, we introduce a publicly available emulation platform built on top of Mininet and the Quagga routing software, for experimenting on hybrid BGP–SDN AS-level networks. As a proof of concept, we focus specifically on exploiting multi-domain centralization to improve BGP's slow convergence. We build and make publicly available a first multi-AS controller tailored to this use case and demonstrate experimentally that SDN centralization helps to linearly reduce BGP convergence times and churn rates with expanding SDN deployments.

## 1. Introduction

BGP has been the de facto inter-domain routing standard in the Internet for the last three decades [1]. Its primary function is the exchange of IP prefix reachability information between Autonomous Systems (AS). BGP routing converges in a distributed fashion into policy-compliant paths crossing the Internet on the AS level. AS administrators apply multifarious BGP policies [2], ranging from basic shortest-path routing to complex traffic engineering schemes for security, cost reduction, and conformance to business agreements, such as customer-to-provider or peer-to-peer relationships [3].

Previous research has thoroughly analyzed the various problems of BGP [4]. One example is its slow and "chatty" convergence [5,6]; one important cause of this behavior is the path exploration problem [7]. Best practices refined over years of operational experience have introduced mechanisms such as Minimum Route Advertisement Interval (MRAI) timers to rate-limit routing advertisements, or selective route flap damping to absorb routing oscillations caused by unstable prefixes [8]. In reality, operators follow diverse practical strategies to achieve fast convergence, e.g., ditching the MRAI usage altogether [9] to avoid "idle waiting" during route propagation, or abolishing route flap damping as harmful [10]. Despite these shifts and changes over the years, the basic problem persists: BGP can take 10s of seconds up to some minutes to converge after a routing change [6]; these times might not be acceptable for the operation of multiple Internet applications, such as Voice-over-IP (VoIP) [11].

* Corresponding author. Tel.: +41762888094.
  *E-mail addresses:* vasileios.kotronis@tik.ee.ethz.ch, biece89@gmail.com (V. Kotronis), gaadrian@student.ethz.ch (A. Gämperli), fontas@ics.forth.gr (X. Dimitropoulos).

Despite the fact that a large number of proposals aiming towards solving the problems of BGP—including slow convergence—have been formed, most of these proposals never leave the research stage. The reasons are multiple. First, BGP is a widely adopted protocol implemented by many stakeholders and is therefore very difficult to change. Second, ISPs cannot be easily convinced to take the risk of adopting a proposed improvement unless substantial profit is imminent. Third, the strict requirement of maintaining backwards compatibility nips many good ideas in the bud; green-field approaches are usually discarded as "utopian". We cannot *simply change* BGP at one go.

In this context, the Software Defined Networking (SDN) [12] architecture offers new opportunities. The key concept of SDN is the separation of the network control plane from the data plane, e.g., separating routing from routers [13]. SDN enables logically centralized Network Operating Systems (NOS) [14–16] and controllers. A NOS implements the state distribution abstraction of the layered SDN model and interacts with packet forwarding elements based on forwarding abstractions such as OpenFlow [12], i.e., the *southbound* interface. Control features and applications, including routing algorithms, can be deployed on top of the NOS and run as software modules using a specification abstraction API, i.e., the *northbound* interface. The NOS presents a consistent network-wide view to the centralized control logic running on top of it. Multiple NOS systems and network applications can run over the same substrate, using network hypervisors implementing the virtualization layer of the SDN control stack [17]. The following question arises naturally in this context: *can we take advantage of SDN concepts, such as logical centralization, on an inter-domain level?*

In this work, we leverage SDN to improve inter-domain routing properties while also enabling innovation in routing applications running across domains. This can be achieved by gradually forming logically centralized inter-domain routing controllers and AS clusters which are served by these controllers. As a financial and technical means towards inter-domain centralization, which is an unconventional idea, we propose to outsource the routing control plane of an AS to external trusted providers, i.e., "Routing-as-a-Service" contractors [18], according to our previous work [19]. The contractor specializes in routing management and can relieve the ASes of the burden of maintaining expensive, highly-trained staff who manage the cumbersome routing complexity [20]. Since a contractor manages routing for several ASes, it can take advantage of this multi-AS level of logical centralization and aggregation in order to improve multiple aspects of inter-domain routing, while maintaining legacy compatibility with non-client ASes. We note that each AS preserves its policy-shaping capability, privacy and business identity; the contractor can for example operate only on a virtual slice [17] of the client network, managing inter-domain interactions. Outsourcing is only a means to an end; there can be alternative paths to inter-domain centralized control, e.g., based on ISP coalitions occurring at Internet eXchange Points (IXPs), mediated via an SDN controller [21].

The contributions of this work are the following. First, we propose a rather radical idea involving a *new routing model*, which is based on *inter-domain centralization*. Routing outsourcing is one way to achieve it [19]; we further present incentives and limitations associated with the model. Secondly, we develop a *publicly available emulation framework* for conducting *hybrid BGP–SDN* inter-domain routing experiments; this can be used in generic experiments also by other researchers. Thirdly, we design and implement a *proof-of-concept SDN controller* which controls AS clusters via OpenFlow and maintains legacy compatibility with BGP. Insights on the development of such a controller are analyzed in detail. Finally, as a use case, we evaluate the *interplay between path-vector BGP and link-state SDN routing* in terms of convergence using the developed controller and framework. Our findings indicate that convergence times can be linearly reduced with increasing SDN penetration in hybrid multi-domain networks, while churn rate reductions need relatively large SDN deployments to be tangible. The experimental results can be replicated by other researchers for verification purposes, as the software and scripts used are available to the community [22].

The rest of the paper is structured as follows. Section 2 provides some background on BGP, SDN and related work in the fields in which our work is applicable. Section 3 describes the inter-domain routing centralization model that we advocate. Section 4 gives an overview of the hybrid BGP–SDN framework on which we run our experiments. Section 5 describes the design goals and implementation insights gained during the development of a proof-of-concept inter-domain routing controller. Section 6 presents the results from the evaluation of routing convergence on hybrid BGP–SDN multi-AS networks. We conclude in Section 7, discussing open questions and future work.

## 2. Background and related work

**BGP path selection.** BGP is a path-vector routing protocol in which every router decides locally the "best" AS path per destination prefix. This choice is based on local policies, AS path lengths, and other attributes, e.g., involved in tie-breakers. The *local preference* attribute is used to set policies for outbound traffic; these policies may correspond to business relationships [3] or day-to-day ISP operations [2]. Filtering, applied on the BGP updates received from or exported to peers, is also a common practice for enforcing policies. BGP routers use MRAI timers in order to rate-limit BGP updates to peers and achieve more stable routing. The MRAI timer is applied per (destination, peer) tuple [1]; the default value in today's Cisco routers is 30 s. Routing loops can be detected and avoided by checking whether a received AS path announcement includes the local AS number. However, state inconsistencies between routers, e.g., during convergence periods, can lead to transient loops [23].

**Inter-domain SDN.** SDX [21] proposes the deployment of SDN-capable data plane elements within IXPs, controlled by an SDN controller; this setup enables richer traffic matching, more direct control over the data plane and new applications such as inbound traffic engineering, WAN load balancing and IXP fabric virtualization. Bennesby et al. [24] propose the use of an inter-AS routing component on per-domain SDN controllers, achieving BGP-like functionality over a distributed AS controller fabric. This architecture enables the decoupling between BGP routing policy and network infrastructure, allowing innovation. As a next step, Thai and de Oliveira [25] introduce an Interdomain Management Layer (IML), based on horizontal slicing of network resources. IML enables independent SDN ASes and provides tools for managing AS borders and sharing resources with other ASes via domain proxies. Lastly, we mention the work on seamless internetworking between SDN and IP [26], and the RouteFlow approach applied in hybrid legacy-SDN networks [27]. In contrast to these approaches, our model is based on logically *centralizing* the routing logic of a *multi-domain* environment.

**Outsourcing network functions.** Sherry et al. [28] propose the outsourcing of enterprise middlebox processing to the cloud. Gibb et al. [29] propose the outsourcing of network functionality to external feature providers. Both studies suggest the export of traffic which is routed in the data plane, while we focus on outsourcing the routing control logic. Lakshminarayanan et al. [18] introduce Routing-as-a-Service, motivated by the resolution of tussles between ISPs and customers over the control of end-to-end paths. Instead, our work focuses on the benefits of partly outsourcing the per-AS routing logic and combining the inputs from multiple ASes for improving inter-domain routing.

**Convergence problems and solutions.** Previous research has studied BGP convergence properties using models of the protocol [5,30,31]. Early experimental studies highlighted the delayed Internet convergence due to BGP [6,32]. A more recent study of BGP dynamics suggests that some aspects are getting better over the years, but temporal artifacts of BGP convergence still persist [33]. The main cause of delayed convergence is the path exploration phenomenon: after a routing change (e.g., due to topological failures) that invalidates a current best path, a BGP router will select a new best path. The router, however, may choose and propagate a path that has been obsoleted during its selection process. This obsolete path may, in turn, be chosen by other nodes as their new best path, resulting in invalid paths being propagated further. Path exploration has been quantified by Oliveira et al. [7]; BGP can actually take several minutes to converge after a routing change. Unstable prefixes may cause persistent route oscillations [8]; the route flap damping counter-measure was one line of defense against such convergence problems, but was later abolished as hurtful for routing performance [10]. The negative impact of slow BGP convergence on VoIP services over the Internet has also been demonstrated [11]. In order to counter the convergence problems of BGP, multiple research proposals have spawned. Bremler et al. [34] propose a modification to the BGP waiting rules in order to limit both the update "chattiness" and the convergence times after link-up events. Lambert et al. [35] propose the addition of a timer mechanism to enforce order in routing messages, reduce path exploration and control convergence time. Godfrey et al. [36] propose a modification in the route selection process of BGP, favoring stability with some deviation from the operator's preferred routes. In BGP–RCN [37], each update message carries information about the specific cause which triggered it; nodes can thus discard new paths that have been obsoleted by the same failure. *Path exploration damping* is analyzed by Huston et al. [38]; its goal is to reduce update churn and decrease average times to restore reachability, as compared to current BGP mechanisms (MRAI). The common denominator of such approaches is the requirement of global modifications to the protocol itself. In contrast, we propose a model that is compatible with BGP and can help improve its behavior through staged deployment.

**Hybrid routing.** Finally, we have seen proposals of new inter-domain routing protocols with better properties than BGP, such as HLP [39], involving link-state routing within the customer cones of tier-1 ISPs and path-vector routing between tier-1s. Alim and Griffin [40] decompose the algebraic specification of a path problem into sub-problems where different protocols are applied. The authors attempt to clarify the trade-offs between fast convergence of link-state and low space requirements of path-vector; however, modeling mixed BGP-like protocols such as HLP [39] is still an open problem due to the inability to adequately model BGP with semirings. In contrast, we emulate a hybrid path-vector and link-state multi-AS environment using production BGP and SDN code, and we measure their interplay regarding convergence as a use case.

## 3. SDN-based inter-domain routing centralization: benefits and challenges

*What new possibilities does inter-domain SDN centralization enable? Could the radical model of routing outsourcing across domains be realized in some form? Which are the main entities of such a framework and how do they interact with each other?* In this section we analyze the associated trade-offs, starting from the promises of SDN within a domain, and continue with the benefits and challenges that our model entails.

### 3.1. Centralizing routing within an AS

The separation of the network control from the data plane and the consequent logical centralization of routing control promises to drastically simplify routing management within an AS [13,16], and provide faster intra-domain routing convergence [41]. Operators can centrally express [42], enforce and check routing policies using the global view that the NOS provides; these policies can be dynamically compiled and deployed [43]. Moreover, if logical centralization and state distribution are performed with control plane resiliency in mind [15], an AS can benefit from scalable routing while lowering the overall management complexity [20]. Besides, having a central AS NOS simplifies the modification of routing applications, as this process can now be achieved solely based on custom software. Today, the control plane on the routers is extremely complex and is comprised of multiple distributed network functions and protocols (OSPF, LDP, RSVP-TE, iBGP,

MP-BGP, etc.), each with their own state distribution mechanisms. SDN can help build simpler control planes by centralizing instead of replicating complexity everywhere. Thus we can have an evolving routing system for the respective AS since intra-domain routing protocols can change easier. It is also simpler to clone the control plane to safely deploy configuration changes, or tune its redundancy properties.

### 3.2. Inter-domain SDN centralization

Assume that we have formed a logically centralized control plane running routing-related processes within an AS. We propose to exploit the benefits of centralization beyond AS boundaries, using a multi-domain NOS that controls a cluster of ASes. With the term cluster, we mean a group of ASes which are served through the same NOS, regardless of whether they have specific bilateral agreements with each other; therefore clusters can be either contiguous or disjoint. One of the most interesting aspects of a multi-AS NOS is that as more ASes choose to use it for cross-domain routing, larger AS clusters are gradually formed. The advantages of the approach grow as the size of the clusters increases (horizontal scaling). Centralizing the routing control logic of many ASes benefits inter-domain routing in multiple ways as we explain below.

**Bird's eye view over multiple ASes.** The central NOS is aware of (parts of) the policies, topologies, and monitoring information of the ASes within the cluster it controls. It is therefore the natural point at which inter-domain policy conflicts and problems can be spotted and resolved, and routing paths can be optimized. Coordination beyond AS boundaries can yield efficient paths even if ASes have different policies and optimization criteria. This helps to improve routing stability and mitigate path inflation [44]. This benefits multiple ASes, even when they are not part of the cluster, as it may result in shorter and more stable end-to-end paths, thus reducing network load on a larger scale. Even in the case where the ASes within a cluster are not adjacent, the global view of the NOS is still important for routing optimization. An example is the establishment of inter-domain end-to-end paths with specific attributes, such as latency. Moreover, if detailed monitoring data are also exported to the multi-AS NOS, security and network troubleshooting can be further enhanced. For example, the NOS may pinpoint the source of a routing anomaly or failure by analyzing the information acquired by multiple parties, correlating it with external sources for cross-validation. Such benefits can only be leveraged when aggregating information from many ASes, including detection of prefix hijacking [45] or DDoS counter-measures [46].

**Inter-domain routing evolution.** Based on the multi-AS NOS, new inter-domain routing algorithms and protocols can be adopted between the members of a cluster. Innovation inside the clusters can be accelerated, while legacy interfaces with the rest of the Internet (BGP) guarantee proper interoperability. For example, we can have lower convergence times and also decreased churn through centrally controlling the dynamics of intra-cluster routing, as we will show in Section 6, or new services, discussed in Section 7. Additionally, hierarchical routing, which benefits routing scalability [39], is enabled at the inter-AS level thus allowing hierarchical routing schemes to flourish along the NOS control chain.

New BGP-like protocols can be defined between NOSes. Rethinking BGP in the context of the communication between NOSes which control multiple ASes is one possible avenue [24]; this can lead to new routing paradigms.

**Challenges.** Forming such a centralized NOS, controlling the inter-domain routing logic of multiple ASes, comes with its set of challenges. We discuss here the technical rather than the financial/political challenges; the latter ones will be analyzed later under the prism of routing outsourcing. First, we need to have backup fail-over schemes in order to keep everything operational even if the AS-NOS communication fails. This can be achieved with NOS agents within the AS that can "think" locally and act when the global NOS is not available [19]. In general, we need a redundant architecture that provides resiliency in case of failures; NOS hierarchies can be a good direction for that purpose [15]. Hierarchical approaches could also be beneficial for scaling up the multi-AS NOS, while tuning the associated state distribution trade-offs [47]. Second, security and privacy for the communication between the served ASes and the NOS should be guaranteed. In addition, proper northbound APIs for inter-domain services should be provided. We note that the challenges associated with scaling up and securing SDN NOS systems are cutting-edge research topics for the SDN community.

### 3.3. Outsourcing routing functionality

**Why outsourcing?** The art of routing encompasses many more skills than the mere knowledge of how BGP or other routing protocols work. This includes the optimization of traffic flows via traffic engineering, correctly mapping Service Level Agreements (SLAs) to policies, coping with misconfiguration and scalability issues, while at the same time properly securing the network. Each of these oftentimes competing goals requires tuning several knobs in the routing protocol(s). Optimizing how packets are routed within an ISP, in order to satisfy numerous operational and economic objectives, is a difficult research problem. Although a number of advanced traffic engineering techniques have been proposed in the research literature, for example based on integer-programming and multi-commodity flow optimization [48], operators in practice may not have the required knowledge at hand to optimize their network utilization through advanced traffic engineering or to improve security, e.g., through deploying sBGP [49]. Operators are often satisfied with a network that is just running. In addition, the router configuration code an ISP needs to develop, debug, and update is extensive, while the manual configuration of routers requires many administrator work-hours and is an error-prone process; routing misconfigurations are common and can be very costly.

**Technical benefits and transition roadmap.** To address these problems, we propose that the routing control logic of a network could be outsourced to a contractor that specializes in routing management, including routing optimization, configuration, troubleshooting, and monitoring. This would constitute a new type of business relationship for technical routing optimizations. The contractor has extensive knowledge on routing and can therefore provide best routing policies tailored to the requirements of a client. Intelligent routing policies and optimizations enable to improve the reliability, performance, and security of a network. From the perspective of a client AS, outsourcing enables it to benefit from advanced traffic engineering, consulting about best routing practices, policy reconciliation, and network troubleshooting. The transition from today's domain-specific situation to an outsourced routing scheme can be handled smoothly by taking several steps. In a first stage, the contractor consults the client about best practices and together they arrive at to a policy plan that satisfies the requirements of the client. Based on the agreed plan, the contractor takes over the handling of traffic and optimizes routing within the client's domain. For networks without intelligent traffic handling, this can yield direct performance benefits, e.g., in terms of network load or other performance metrics. Besides, the contractor monitors how network traffic in the client's network changes over time and enforces corrective traffic engineering actions.

**Interaction with diverse clients.** Since routing is part of the main business of an ISP, it can choose to outsource only specific parts of routing functionality, e.g., only inter-domain routing without revealing any internal topology information, based on network slicing mechanisms [17]. From the perspective of a client enterprise edge network, outsourcing can help offload the heavy tasks of routing (which is not the primary business of the client anyway) to a specialized party under a contract. Technically, the routing logic of the service contractor calculates the proper configuration of the control plane, updates the state of the network elements of the client AS, and deals with inter-domain routing with the rest of the Internet through BGP. In particular, the client can choose to export the following information to the contractor.

*Routing policies.* These are policies of the client defined by the client AS administrators, or derived based on requirements of SLAs between the client and other parties. They should be enforced and monitored by the contractor. Routing outsourcing does not impede other services offered by a client AS that may depend on routing. This is because the enforced policies are specified by the client AS during the consulting phase and take the requirements of all offered services into account. In addition, the client may regularly update its routing requirements in a dynamic service environment.

*Network's state and monitoring data.* The client exports selected topology, configuration, and measurement data, e.g., network utilization or bandwidth allocation. The contractor is a trusted third party that treats this data as well as routing policies confidentially. The model of a trusted third party, although it requires trust, has been very successful in practice for many modern services. Also, SLAs can always specify the level of confidentiality and traffic visibility, while virtualization and slicing mechanisms [17] implement the needed abstractions from a practical point of view.

*eBGP sessions.* The contractor handles the eBGP sessions and routing interactions between the client and other ASes. BGP messages can be redirected from the border gateways of the client AS to a the contractor's routing control platform and vice versa.

**Financial benefits.** Operators have traditionally viewed the network as their core business. However, declining profit margins have put them under pressure to reduce costs and to launch new, higher margin services. This situation has also pushed operators to streamline their operating expenses (OPEX) and has given rise to an emerging market of managed services, in which the operation and maintenance of the network is outsourced to a third party. In this context, we propose a new model of network outsourcing, i.e., routing outsourcing, which enables the logical centralization of the routing control plane beyond AS boundaries. Financially, the contractor enjoys an opportunity for an economy of scale, as the basic principles of routing optimization are the same across different networks. Economies of scale have been prolific in many computing contexts [50]. We claim that this also holds for routing management. Also, outsourcing can reduce network-related OPEX for the client, via streamlining. Lastly, outsourcing a low-margin service enables more effective use of human resources on higher priority services.

**Challenges.** The transition from current network setups to outsourcing-enabled environments is a multi-stage process; one challenge is the capability to backtrack or change to a better-suited contractor during this process. Therefore, in each of these stages, the client should be able to scrutinize the effects of the changes performed, both in terms of traffic management and expenses, and step back in case it is not satisfied. On another note, policy conflicts between ASes can lead to tussles, which can create problematic paths or even depeering events. The bird's eye view enables the contractor to efficiently detect tussles [51] between its clients. The job of the contractor is to allow the tussles to unfold as today, based on the choices and policy requirements of each client. The main difference is that the contractor can detect and mediate the resolution of routing problems, which may stem from these tussles. In addition, the bird's eye view of the contractor can help find better solutions that meet the policies of each AS than when ASes act alone based on their limited local view. Moreover, as the contractors start competing for clients, additional tussle dimensions arise, thus enabling a new game [51] between the outsourcing entities; this game needs to be further investigated.

In a nut-shell, we propose a scheme where a contractor can control the inter-domain routing logic of multiple ASes based on their policy requirements and network state; as more and more ASes choose the same contractor, AS clusters are gradually formed. These clusters are the manifestation of gradual inter-domain routing centralization, and give us the footing to research the interplay between them and the rest of the Internet. Moreover, we envision multiple contractors competing for clients, and interfacing with each other over new APIs; a full overview of the routing model is given in Fig. 1a.

## 4. SIREN: a hybrid BGP–SDN emulation framework

SIREN [22,52] is a publicly-available Python-based network emulation framework for conducting hybrid BGP–SDN experiments. It extends the Mininet emulator [53], which is a popular environment for SDN experiments. Mininet offers OS-level virtualization (based on Linux namespaces), which efficiently scales up to dozens of emulated nodes and links, and comes bundled with the OpenVSwitch virtual OpenFlow switch [54]. In SIREN, we combine Mininet with the popular Quagga routing software [55], which implements BGP and also other routing protocols such as RIP and OSPF. In Fig. 1b we show the components of a sample SIREN setup. On the left side, we see the legacy BGP part of the emulated multi-AS network, whereas on the right side we illustrate an SDN cluster, composed of OpenFlow switches. BGP routers and SDN switches can originate prefixes. It is also possible to add hosts with IP addresses within a particular prefix for monitoring end-to-end connectivity with tools like ping, etc. All BGP routers peer with a BGP route collector, which collects routing updates for monitoring purposes. Moreover, within the SDN cluster we have a special BGP speaker, called *cluster BGP speaker*, which relays routing information between external BGP routers and the SDN controller. This speaker is implemented with ExaBGP [56]. For every BGP peering there is a link from the speaker to the border SDN switch, in order to relay control plane information over the switches (i.e., eBGP session "outsourcing").

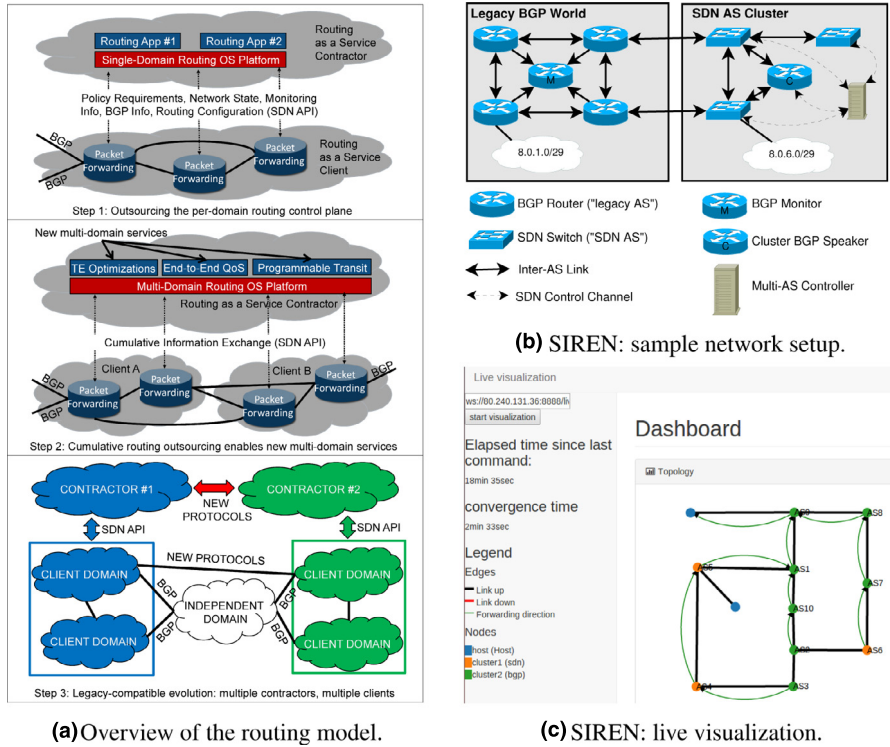

**Fig. 1.** Routing model, steps towards inter-domain centralization (a) and SIREN overview (b,c).

In SIREN, every AS is abstracted as one "big switch/router", i.e., it is emulated by a single network device/node. This abstraction is not fundamental to the framework (and can be extended in future versions), but is useful for use cases such as ours, for the following reasons. First, we want to isolate the effect of *inter-domain* routing convergence and experimentally study its properties. Second, we assume that each AS is not willing to share its internal topology for privacy reasons with the routing outsourcing contractor's controller. The "big switch/router" view is suitable for inter-domain routing management, assuming that the AS is consistent regarding its interactions with other domains at different peering points. This appears to be true in practice [9]. A legacy AS is modeled by a Quagga BGP router, while a SDN cluster AS is modeled by an OpenFlow switch. A cluster AS can use non-SDN mechanisms for internal routing; this does not hinder the view of the AS as a centrally controlled "switch" or "router".

Experimental setups can be written in Python. The framework automatically assigns IP addresses and configures network devices using pre-defined templates. We extended Mininet with several BGP-specific commands to announce prefixes, wait until BGP has converged, etc. Additionally, the framework supports tools for automatic log file analysis, network graph creation, convergence time and loss measurement, and route change visualization. For example, to facilitate experiments on routing stability, the framework detects when the network has converged and whether there is stable connectivity between all hosts. Other compatible tools can be added as Mininet is an extensible platform. Also, experiment batches can be distributed over multiple computing nodes using the *experiment manager*. An example of live routing visualization is presented in Fig. 1c. Forwarding is tracked towards the depicted hosts based on the routing configuration of the ASes on the end-to-end path. The user can visually interact with SIREN via bringing inter-AS links up or down, actively creating convergence triggers and monitoring the network's response. The SIREN framework has been demonstrated at SIGCOMM [52].

MiniNext [57] is another hybrid SDN-legacy routing emulator based on Mininet and Quagga. However, while MiniNext aims at emulating operational environments and focuses on low-level APIs, our framework focuses on multi-AS inter-domain experiments for research and provides a high-level API for experiment orchestration.

## 5. Multi-AS routing controller

In this section, we describe the design and implementation of a first multi-AS routing controller, tailored to improving BGP's slow convergence. We make the controller public as part of the SIREN software [22]. Our goal is not to build a general-purpose multi-AS NOS—the full set of challenges in building such a system is beyond the scope of this paper. In contrast, our objective is to improve BGP's problematic convergence as a proof of concept of our inter-domain SDN model. We use the controller to evaluate the interplay of centralized routing within an SDN-enabled AS cluster and distributed path-vector BGP routing outside of the cluster in Section 6, focusing on convergence.

*5.1. Design goals*

**Exploit centralization.** We wish to exploit routing centralization on the AS level to improve BGP's convergence time and reduce routing churn, leading to more stable routing overall. This helps both ASes in the cluster and the outside (legacy) world.

**Interoperation with BGP.** As BGP is currently deployed across the globe enabling $\sim$ 45,000 ASes to mutually exchange routing information, it is crucial that the controller is fully compatible with the BGP standard [1]. In addition, the SDN cluster should be transparent to the outside world, i.e., legacy BGP routers should think that they talk to yet another BGP router, rather than a multi-AS SDN controller.

**No cluster lock-in.** The identity of the participating ASes (e.g., their AS numbers) should be preserved in the hybrid routing system. This prevents cluster lock-in, which could result if AS numbers were replaced with a "super-domain" identifier. Instead of this, we wish to have groups of ASes which are visible as separate entities to the rest of the Internet and maintain their individual identities and policies. This approach also facilitates a smooth transition to the new system as existing mechanisms relying on AS numbers (e.g., access lists or BGP communities) do not need to be updated.

**Disjoint clusters.** As the transition to the new architecture will likely be gradual, clusters will probably not be contiguous, at least in the beginning. This means that AS paths may enter, exit and reenter the cluster at different points (e.g., IXP-facing ports); thus the controller should be able to calculate paths using the global view of a disjoint cluster of clients and the legacy BGP information that it learns through them. This also means that in case the cluster is internally partitioned due to an inter-domain link failure, it may not be partitioned on the global level since paths that join the two parts over legacy ASes can still be used. Thus reachability over disjoint clusters is achieved.

**Hybrid routing.** The controller knows the full topology within the cluster and receives external AS path announcements from the outside world via BGP. Therefore inter-domain routing becomes hybrid path-vector and link-state [40]. The controller can use an algorithm such as Dijkstra in order to calculate shortest paths over the cluster topology. External AS paths learned from BGP can be attached as "extensions" to the cluster graph and be explored with Dijkstra. Selected paths can be then advertised to legacy BGP peers, making the controller a part of the outside BGP path-vector system.

**No loops.** AS-level loop avoidance is essential in the new hybrid BGP–SDN setup both for routing efficiency and correctness. We note that naively using the same loop avoidance mechanism as BGP is not wise, as we will show later, due to the differences between BGP's distributed local view and SDN's global view approach.

*5.2. Implementation details*

The controller runs using POX [58] mechanisms for OpenFlow-based interaction with the cluster switches, and interfaces with external BGP routers through ExaBGP [56]. POX-like cooperative multitasking is used for the event-based processing that happens on the controller. This approach is well-suited for rapid prototyping; we can focus more on research questions rather than state consistency, scale and concurrency issues [15]. To better understand the operation of the implemented path selection algorithm, we first introduce two graphs representing the core state that the controller maintains.

**Switch Graph.** The *Switch Graph* is a simple directed graph that represents the physical topology of the cluster combined with prefix connectivity information, as seen from the controller's perspective. We have two kinds of nodes: switch nodes, which represent SDN switches, and prefix nodes. The presence of an edge means that data can be forwarded from the source to the destination of the edge (prefix to switch, switch to switch). The *Switch Graph* is built gradually: we add a directed edge between two switch nodes, when a switch node detects a link in that direction. An edge from a switch to a prefix node is added, when the prefix is learned from BGP or the prefix is directly connected to that particular SDN switch. In the "BGP-learned" edge case, we only add the best path in terms of hop count and annotate the edge with the corresponding AS sequence. We save all paths which the cluster receives information about; best (i.e., shortest) paths are then selected for the eventual routing of traffic across the ASes.

**AS graph and loop avoidance.** In our hybrid link-state / path-vector setting, we need to cater for AS-level paths that leave and re-enter the cluster. If such paths were naively marked as annotations to external prefixes as in the *Switch Graph* and used directly by Dijkstra, then we could get loops. For this purpose, we break such paths into two parts: (1) a destination prefix attached to the last cluster AS in the path; and (2) virtual links that connect cluster ASes over external paths. We incorporate these changes into a per-prefix *AS Graph* structure, which is a transformation of the *Switch Graph*. At the beginning of the transformation, all AS numbers of the cluster are added as nodes. The AS connections inside the cluster, which have been represented as edges between switch nodes in the *Switch Graph*, are also added to the new graph. The transformation is therefore restructuring the *Switch Graph* taking into account paths that cross the legacy world and the SDN cluster in order to avoid loops. Fig. 2a shows an example *Switch Graph*. Switches 1–3 form a cluster. Switches 1 and 2 know a path to prefix 8.0.10.0/29, which they learned over BGP; these paths pass over external legacy ASes. Note that the path known to switch 2 passes over

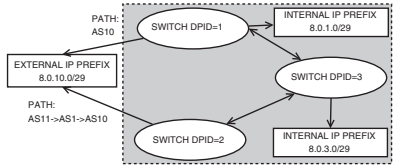 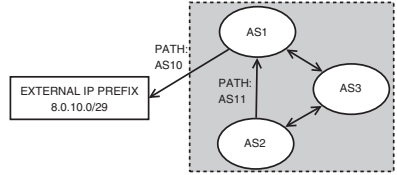

**(a)** Example of a Switch Graph.    **(b)** Example of an AS Graph.

**Fig. 2.** Example of Switch-to-AS graph transformation: paths to 8.0.0.10.0/29 are sanitized to avoid loops.

switch 1 as an intermediate node. Switch 1 has a directly connected prefix, and so does switch 3. Fig. 2b shows the derived *AS graph*: the switches have been transformed to their corresponding ASes. Note that the path known to switch/AS 2 has been sanitized; we have added a virtual link that includes the external AS path that exits and re-enters the cluster (in this case over AS 11), ending at AS 1. AS 1 then knows the best (shortest) path towards the prefix. This process guarantees that if BGP does not induce loops until this AS, then shortest path routing on the controller's side will result in AS-level loop-free paths.

**Main algorithm.** Dijkstra is run on the AS topology graph using AS path weights. This allows the controller to calculate the shortest paths towards each prefix learned either internally from the cluster ASes or externally over BGP; paths to external prefixes exiting and re-entering the cluster are sanitized for loop avoidance as explained before.

**Path recomputation problem.** Paths are only recomputed when needed. A link change between switch nodes or a switch change in the *Switch Graph* results in a full recomputation of the paths associated with all prefixes currently known in the network. However, when only a path to a certain prefix is changed, e.g., when a second switch adds a new path to a specific prefix (learned over BGP), only paths leading to that single prefix will be recomputed. At this point we should note the following insight gained during the implementation process. The SDN cluster controller can receive multiple BGP updates per second stemming from external ASes, since it controls the inter-domain routing interactions of several ASes, each one with multiple external peers. Each of the updates triggers changes in the switch and AS graphs, causing path recomputation throughout the cluster. We need to stress out that this is an expensive process; path recomputation is equivalent to switch reconfiguration through manipulation of the flow tables. Installing all necessary rules on the associated switches can take 100s of milliseconds; during this time more BGP updates are received stressing the process even further. Moreover, the controller's actions need to be advertised to external peers; that means that besides the traffic shifts caused inside the cluster due to the flow rule installation process, the instability will also propagate further outside of the cluster and cause further problems.

**Delayed path recomputation.** In order to mitigate this issue we added a mechanism for delayed recomputation of paths, based on a timeout value called "Cluster Waiting Recomputation Interval" (CRWI). This is different than the MRAI advertisement interval of BGP. After the CRWI timeout happens, we compute and install locally the rules associated with the new paths via OpenFlow. These paths are the result of queued recomputation requests, accumulated over the waiting interval; we then directly advertise the changes over BGP to the outside world. This strategy helps us avoid routing inconsistencies with neighbors due to outdated information, since the queued requests are sanitized in terms of age. Furthermore, it can help make the network more stable by "rate-limiting" the cluster controller, reducing the number of required path changes and leaving some temporal slack for the forwarding rules to be installed on the cluster switches. In our experiments, we found that a CRWI of 1 s is sufficient to avoid any problems with routing inconsistencies and flow rule installation delays.

**Other details.** The controller has partial support for consistent state updates during the reconfiguration of the cluster switches. Proxied control traffic (BGP) and direct data (ARP, IP) traffic are both handled via flow rules. The controller and its operational features (e.g., topology detection) have been demonstrated at SIGCOMM [52].

## 6. Evaluation of routing convergence

### 6.1. Experimental setup

*What is the effect of inter-domain SDN centralization on BGP convergence time and stability?* As a proof of concept of our routing model, we evaluate the effect of SDN centralization using our multi-AS controller and SIREN. In our experiments, a dual-homed AS loses its primary connection and fails-over to its backup link. Its two providers are selected at random from a set of ISPs connected to each other in diverse topologies. To enforce the primary-backup setup, the client AS prepends its AS number multiple times in its prefix announcements propagated over the backup link. The link-down event on the primary link causes a wave of withdrawals throughout the network, accompanied with announcements of new—but not always valid—paths due to the path exploration process. The ISPs explore alternative paths to the client, taking into account the prepended route advertisements going over the backup—now active—link, as they converge to the shortest path. In this setting, we evaluate how gradual SDN penetration, in terms of increasing percentage of cluster SDN ASes, affects the convergence time and the average routing update churn rate. We note that we are using fast keep-alive and hold-down timers for both Quagga and ExaBGP, since we want to explore what happens *after* the link-down detection and not waste time discovering that the link is down. The timer values are selected in a way that avoids negative synchronization effects.

**Simplifications.** We make the following simplifications in order to fit our use case and the properties of the data we have at our disposal (i.e., AS-level graphs). First, we assume one node per SDN-controlled AS ("big switch/router" approach) as already described beforehand. We understand that this is an important simplification [59], but it allows us to capture some basic properties of hybrid inter-domain routing even without knowing how an AS is structured internally. Also, according to the survey from Gill et al. [9], ASes are usually consistent regarding their routing information export across their distributed fabric. Second, we assume that where we do not have explicit policies, the controller calculates Dijkstra-based shortest paths. In BGP terminology, we are taking into account the AS path length. We note that while there are ways of running centralized Dijkstra to find policy-compliant shortest paths [60], obeying the Gao-Rexford conditions [3], inter-domain policies are in reality much richer and more diverse than that. The main problem is that they are quite difficult to infer and are by nature commercial secrets of the ISPs; this results in lack of data regarding what policies ISPs actually implement, leaving only qualitative surveys [2] or surveys on small AS set samples [9] to extract information from. Therefore, to simplify our experiments, we chose to explore shortest path dynamics ignoring complex policies.

**Table 1**
Parameters used for the inter-domain routing convergence experiments.

| Setup parameters | Values |
| --- | --- |
| Experiment type | route fail-over for dual-homed client |
| Topology type | Clique (full mesh), Erdos–Renyi (E–R) [62], Barabasi–Albert (B-A) [62], Newman–Watts–Strogatz (N–W–S) [62] |
| Topology size [number of nodes] | 8, 16, 32 |
| Topology size [number of links] | function(node_number, graph_type) |
| Clusters and controllers [number] | one controller, one (contiguous or disjoint) cluster |
| SDN penetration [%] | 0, 25, 50, 75 |
| SDN cluster CRWI [second] | 1 |
| BGP MRAI on Quagga [second] | 0, 30 |
| Keep-alive timer on Quagga, ExaBGP [second] | 5 |
| Hold-down timer on Quagga, ExaBGP [second] | 15 |
| Reconnect timer on Quagga, ExaBGP [second] | 5 |
| Policy for BGP ASes, SDN ASes | prefer shortest AS path (hop count) |
| Client policy for backup link | use 10-fold ASN path prepending |

**Emulated topologies.** Regarding AS-level topology emulation, we initially considered the CAIDA *IPv4 Routed /24 AS Links* dataset [61], providing snapshots of AS links derived from IP-level topology measurements. Due to the large size of the dataset in terms of AS nodes and links (∼tens of thousands), which goes beyond the scalability limits of Mininet, we did not run experiments on these graphs, but used synthetic topology models instead. According to the seminal work of Willinger and Roughan [59], there is not yet a widely accepted model of the AS-level Internet topology; such inference requires a cumbersome reverse engineering approach based on domain-specific knowledge. Therefore we took multiple different models into account [59]: cliques (full meshes), random graphs such as Erdos–Renyi, scale-free graphs of the preferential attachment type based on the Barabasi–Albert model, and small-world graphs using the Newman–Watts–Strogatz approach. We then searched if common patterns were replicated across different graph types and scales, indicating interesting BGP–SDN interactions. We used the NetworkX graph generator [62] and selected its parameters such that the derived graphs constitute a compromise between fully connected ISP meshes, and sparse tiered environments. Larger parameter values are closer to the first setup, while smaller values to the latter. The full set of experimental parameters and values explored is presented in Table 1. The parameterized code and scripts, together with instructions on how to use the SIREN framework, are publicly available [22].

### 6.2. Experimental results, observations & insights

The results from our experiments are depicted in Fig. 3, regarding convergence times, and Fig. 4, regarding churn rates. The results are based on an MRAI of 30 s; our findings using MRAIs of 0 s were very similar and are omitted from the presentation for space reasons. The cause for this similarity is that Quagga route withdrawals are not rate-limited (in contrast to announcements), while at the same time being the main triggers for the path exploration process. This process primarily affects the convergence results that are seen in the figures. Therefore, a first insight we gained was the importance of the withdrawals for path exploration and the indifference that the MRAI value has on the results. Further observations and insights follow.

**How does the scale of the graphs affect convergence, taking into account varying levels of SDN penetration?** Convergence times exhibited a small-gradient linear decrease at the 8-node scale, with comparable times in E–R, B–A and N–W–S graphs for different SDN penetration levels, with the most notable gains in the clique case. At the 16-node scale, we observed a high-gradient linear decrease of convergence time with increasing size of the SDN cluster. Finally, at the 32-node scale, a negative sub-linear relationship between convergence time and SDN penetration is observable. In this case, at 25 and 50% SDN penetration, the reduction in convergence time is slower. However, the convergence time drops rapidly between the 50 and 75% levels, where the time is cut by more than half. We also note the very small width of the boxplot at the 75% SDN penetration cases across all scales and topology types; this indicates very small variance on convergence times due to a stabilizing effect of centralization.

Moreover, the absolute convergence times are effectively doubled as the topology doubles in size. The same rule applies for churn rates; bigger scales translate to higher churn (updates/second). At the 8-node scale, the churn rate ranges

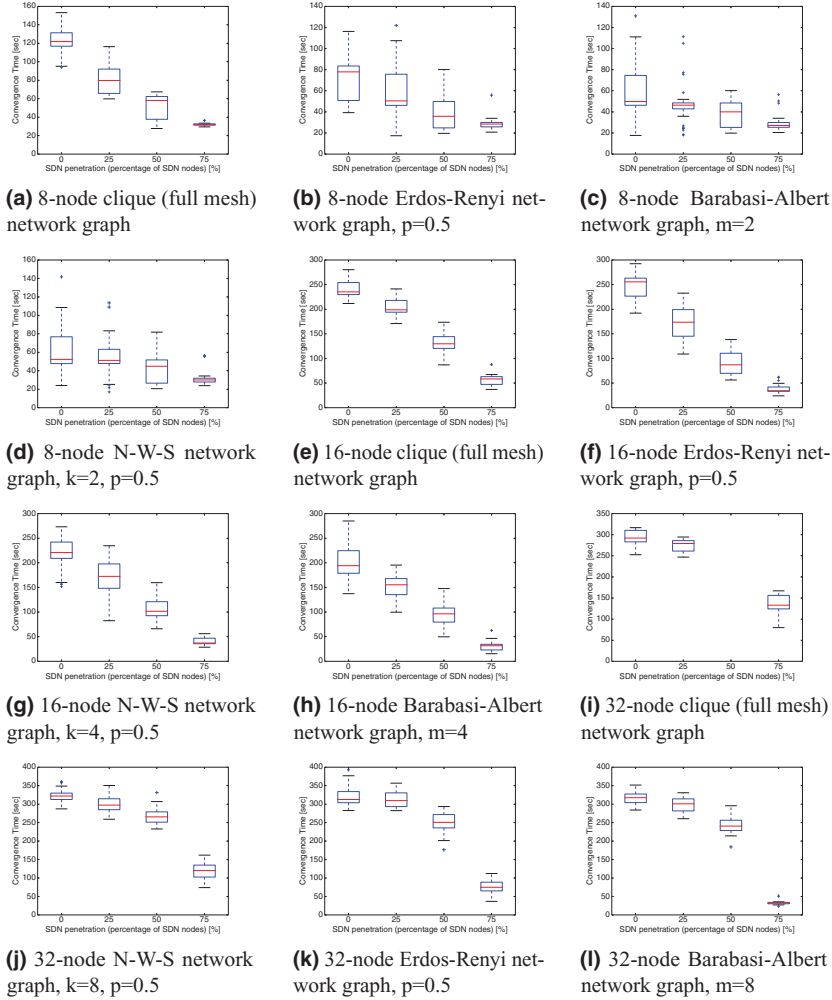

**(a)** 8-node clique (full mesh) network graph

**(b)** 8-node Erdos-Renyi network graph, p=0.5

**(c)** 8-node Barabasi-Albert network graph, m=2

**(d)** 8-node N-W-S network graph, k=2, p=0.5

**(e)** 16-node clique (full mesh) network graph

**(f)** 16-node Erdos-Renyi network graph, p=0.5

**(g)** 16-node N-W-S network graph, k=4, p=0.5

**(h)** 16-node Barabasi-Albert network graph, m=4

**(i)** 32-node clique (full mesh) network graph

**(j)** 32-node N-W-S network graph, k=8, p=0.5

**(k)** 32-node Erdos-Renyi network graph, p=0.5

**(l)** 32-node Barabasi-Albert network graph, m=8

**Fig. 3.** Convergence times vs SDN penetration for the fail-over experiment: emulation results on graphs of different type and size, for BGP MRAI = 30 s. Boxplots correspond to 20 emulation runs. The reader is referred to the NetworkX website [62] regarding graph parameterization (i.e., *k, p, m*).

between only 1 and 3 updates/second without major changes with increasing SDN penetration. At the 16-node scale, the churn rate increases slightly to up to 12 updates/second and SDN penetration shows more clear gains, although it might even slightly increase churn in some cases. Finally, for the 32-node size, the churn rate increases much faster to up to 60 updates/second, and SDN penetration leads to consistent reductions in churn. At this scale the churn exhibits a sublinear decrease with increasing SDN penetration, similarly to the convergence time patterns. This is because while the control plane state is propagated at a higher pace via the controller, the CRWI-based rate-limiting on the controller's side smoothens the convergence process.

**How does the network graph type affect convergence, considering varying levels of SDN penetration?** The behavior of convergence time and churn rates were not significantly affected by the graph type, and the patterns we observed regarding SDN penetration were more or less preserved across diverse topologies. What matters more is the scale of the topology, as already explained beforehand. We note that the clique has only slightly different behavior regarding absolute numbers; in fact convergence times and churn rates were always elevated in contrast to the other topologies. This is expected since the clique is essentially the "worst-case" scenario for BGP convergence; we verified this fact experimentally. Setups that are sparser than the clique also seem to benefit from increasing SDN penetration in similar ways, leading though to faster and less "chatty" convergence due to the less intense path exploration process.

**What is the actual effect of convergence on data plane traffic?** In our experiments we focused primarily on the behavior of the control plane during convergence. Further examination of the interaction between the control and data plane yielded the following insights. *(i)* Delayed convergence primarily affects the latency, jitter and ordering of the data packets; in the fail-over case packets usually travel upon the different explored routes (even in circles) until the final valid paths become available. *(ii)* We observed negligible packet loss during convergence. That means that all intermediate nodes always have fail-over paths towards a destination, meaning that the packets eventually reach it, albeit following different routes until stability is restored. Correlations between packet loss and BGP events in the Internet have been observed in the work of Kushman et al. [11]. Lastly, for statistics on the frequency of instability events and the percentage of affected BGP prefixes, we refer the reader to the work of Huston [63].

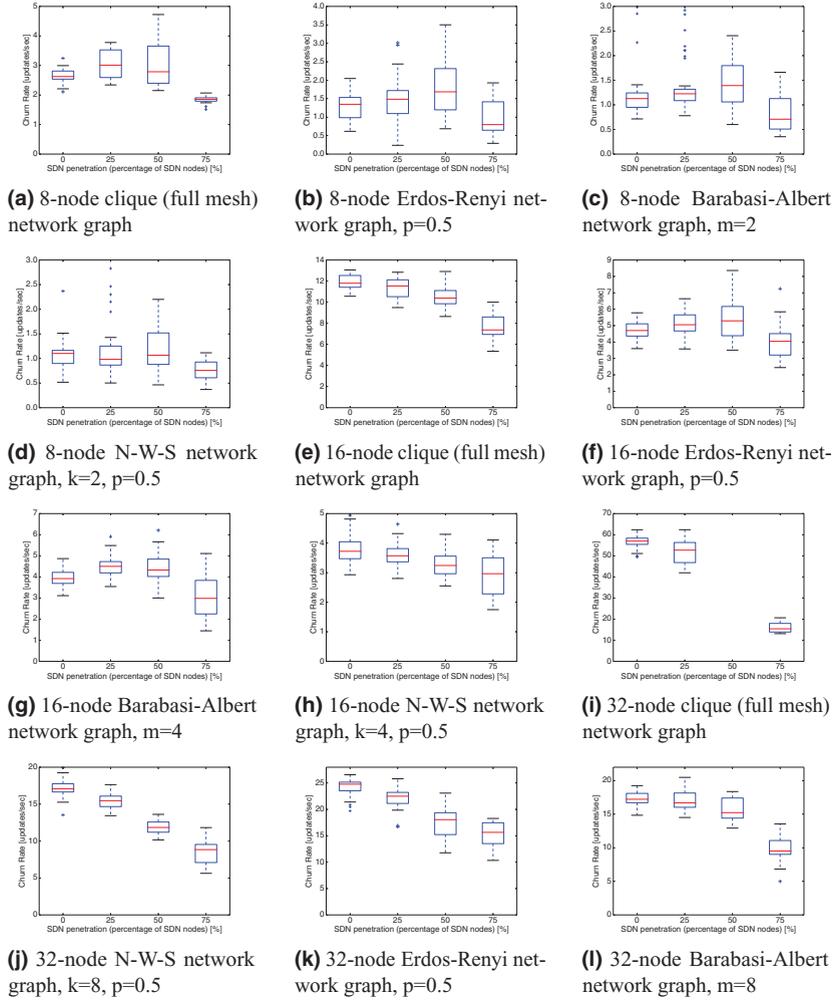

**Fig. 4.** Average routing update churn rates vs SDN penetration for the fail-over experiment: emulation results on graphs of different type and size, for BGP MRAI = 30 s. Boxplots correspond to 20 emulation runs. The reader is referred to the NetworkX website [62] regarding graph parameterization (i.e., *k, p, m*).

**What are the key take-away messages regarding the interplay between legacy BGP AS groups and SDN AS clusters?** Gradual deployment of SDN and inter-domain routing centralization actually helps. Benefits in convergence times can already be seen with small penetration levels, while benefits in churn rates need larger deployments to be tangible. In our experiments, the critical mass that a Routing-as-a-Service contractor should acquire in order to improve the stability of a multi-domain network seems to be somewhere between 25 and 50%. Between these levels and at the 32-node scale, convergence times can be reduced by ∼20%, while churn rates by ∼ 15%. In general, we observed that the use of logical centralization of routing control accelerates convergence because of two main factors. *(i)* The state propagation process is accelerated due to the central point where parts of the state are gathered and are then directly communicated outside. This acceleration benefits both client and non-client ASes, but may increase the associated churn in some cases. *(ii)* The controller has a global overview of its cluster and the inter-domain network overall; this view is efficiently used for informed decisions related to path exploration, based on the cumulative routing feedback.

## 7. Conclusions and future work

**Conclusions.** We proposed the gradual centralization of parts of the routing control logic of multi-domain AS-level networks. The goal is to improve general properties of inter-domain routing, such as the convergence behavior accompanying routing changes. The proposal can be technically applied using SDN mechanisms, while Routing-as-a-Service outsourcing frameworks may offer a financial basis for market adoption. As a use case, we evaluated the interplay between SDN-based routing centralization and classic BGP routing. To support that, we developed a hybrid BGP–SDN emulation framework and a multi-AS SDN controller running on top of it. Our fail-over experiments on hybrid graphs of diverse scales and types indicate that inter-domain routing centralization improves convergence times even at small SDN penetration levels. Churn rates are comparable or slightly worse than pure BGP at small scales, with benefits shown at larger scales. Our work is another step towards extending the value proposition of SDN on the inter-domain level [21,24], based on the radical idea of logically centralizing the inter-AS routing control plane [19]. We note that this is one of the most challenging arenas for SDN to penetrate, due to the difficulty of making changes on how core routing works; politics and established practices can put a brake on novel technical approaches. Nevertheless, our current findings encourage further research along this direction of inter-domain SDN. We highlight the following avenues of future work.

**Abstractions and services.** A multi-domain routing control platform can become a vehicle for the deployment of novel services, which are hard to implement in today's environment. This requires the identification of the proper abstractions that will be offered to the services running over the platform. In this context, we can take advantage of layered control channel architectures using network programming languages [42] and compilers [43], identifying the proper *northbound* interface between the control platform and the multi-domain services. The virtualization/slicing abstraction [17] is another piece of the puzzle. One potential service that would be interesting to run on a cross-domain level using our platform, is collaborative defense against new DDoS attacks, such as the *Crossfire* link-flooding attack [46]. Such a service could for example take advantage of SDN-based traffic engineering for *joint* detection and mitigation.

**Controller trade-offs.** Future work includes the quantification of the scalability, resiliency and centralization trade-offs for the multi-AS cluster SDN controller, based on the lessons learned from the ONOS [26] and ONIX [15] projects. For example, proper controller placement in a multi-domain setting so as to deal with latency and distribution trade-offs is an interesting avenue to explore [64], as well as possible fail-over setups and the resulting state consistency of the control and data planes upon fail-over events.

**Policy support.** Policy support is another aspect of framework extensions, combined with the policy interactions between inter-domain services running on top of the controller. Efficient algorithms for computing policy-compliant shortest paths and path diversity for arbitrary topologies and a variety of policies are part of our ongoing work.

**Taking convergence out of the critical path.** We would further like to explore maximally redundant techniques for fast re-routing on the IP layer, such as the ones that are currently under discussion in IETF [65]. The objective there is to minimally disrupt traffic upon rerouting; as the convergence process itself is less of an issue. Moreover, we plan to investigate complementary IETF efforts on making Internet routing more scalable, such as LISP [66]. LISP can be used to reliably forward traffic to prefixes, even while the network is converging, while being backwards compatible with BGP and reducing the amount of the needed signaling. We note that such mechanisms could be safely deployed within the sphere of influence of a SDN controller, benefiting client ASes while shielding the rest of the Internet from any associated issues.

### Acknowledgments

We would like to thank Dr. Bernhard Ager from ETH Zurich for his useful feedback. This work is partly funded by European Research Council GA no. 338402.

### Supplementary material

Supplementary material associated with this article can be found, in the online version, at 10.1016/j.comnet.2015.07.015

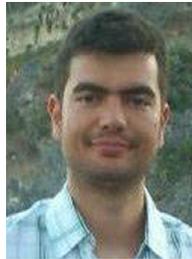

**Vasileios Kotronis** holds a Diploma in Electrical and Computer Engineering (Dipl.-Ing) from the National Technical University of Athens (NTUA, July 2011). He has been a member of the Communication Systems Group at ETH Zurich as a doctoral student since December 2011. He worked on the project OpenFlow in Europe - Linking Infrastructure and Applications (OFELIA-FP7) between December 2011 and October 2013. His main research fields are: Software Defined Networks (SDN), Internet Routing and SDN Testbeds.

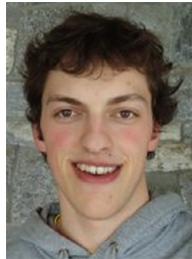

**Adrian Gämperli** holds a Master's Degree in Electrical Engineering and Information Technology from the Swiss Federal Institute of Technology in Zurich (ETH Zurich). His main areas of research are data networks and network security.

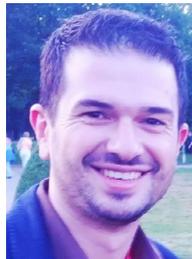

**Xenofontas Dimitropoulos** is an Assistant Professor in the Computer Science Department of the University of Crete, since January 2014. In addition, he is affiliated with the Institute of Computer Science of the Foundation for Research and Technology Hellas (FORTH) and with the Communication Systems Group of ETH Zurich. He leads a research group working on Internet measurements and software defined networks with the goal of making the Internet more reliable and secure. Before, he worked in the Georgia Institute of Technology, the IBM Research Labs and the University of California San Diego (UCSD). He has published 64 papers and 3 patents and has received 2 best paper awards. In addition, he has received prestigious grants from the European Research Council, the Marie Curie and the Fulbright programs. In 2014, he served in the Organizing Committee of the flagship networking conference, ACM SIGCOMM.